\documentclass[a4paper, 10pt, reqno]{amsart}
\usepackage[margin=1in]{geometry}
\usepackage{amsmath}
\usepackage{amssymb}
\usepackage{mathtools}
\usepackage{tikz}
\usepackage{enumerate}
\usepackage{gensymb}
\usepackage{pgfplots, array}
\usepackage[width=\textwidth]{caption}
\usepackage{filecontents}

\title{Empirically Improved Tokuda Gap Sequence in Shellsort}
\author{Ying Wai Lee}
\date{\today}
\begin{document}
\maketitle
\begin{abstract}
Experiments are conducted to improve Tokuda (1992) gap sequence in Shellsort into $\gamma$-sequences, and the best result is the gap sequence in which the $k$-th increment $h_k$ is given by
\begin{align}
h_k=\left\lceil
\frac{\gamma^k-1}{\gamma-1}
\right\rceil
\end{align}
, where $\gamma=2.243609061420001...$ and $k\in\mathbb{N}_1$. The first few increments of the gap sequence are
\begin{align}
1,\, 4,\, 9,\, 20,\, 45,\, 102,\, 230,\, 516,\,1158,\,2599,\,5831,\,13082,\,29351,\,65853,\, 147748,\,331490,\,743735,\, ...\end{align}It empirically yields less numbers of comparison on average than Tokuda (1992) gap sequence. In the procedure of search, it reveals the potential existence of a new type of fractal. \end{abstract}

\section{Introduction}
Shellsort, invented by Shell (1959), is one of the most traditional sorting algorithms, in which a permutation is sorted by several iterations of linear insertion sort on the elements that stay apart by a positive integer from a gap sequence, called an increment that progressively decreases to 1. 

Ciura (2001) gap sequence is empirically the best known gap sequence in the sense of having empirically the least average numbers of comparisons in Shellsort for relatively small number, say 4000, of elements, (Ciura, 2001), which is a gap sequence consisting of only finitely many increments given by (Ciura, 2001),
\begin{align} \label{Ciura}
1,\,4,\,10,\,23,\,57,\,132,\,301,\,701
\end{align}

One of the practical approaches to extend the sequence in (\ref{Ciura}) is to apply (\ref{re2.25}) for which the $k$-th increment $h_k$ is given by, where $k\in\mathbb{N}_9$ and $\lfloor x\rfloor$ denotes the floor function,
\begin{align}\label{re2.25}
h_{k}=\lfloor2.25h_{k-1}\rfloor
\end{align}

The value $\gamma_0=2.25=9/4$ appears in Tokuda (1992) gap sequence, which is the previously best known gap sequence for relatively large number of elements, in which the $k$-th increment $h_k$ is given by (\ref{Tokuda}) (Tokuda, 1992), where $k\in\mathbb{N}_1$ and $\lceil x\rceil$ is the ceiling function.\begin{align}\label{Tokuda}
h_k
&=\left\lceil\frac{1}{5}\left(9\left(\frac{9}{4}\right)^{k-1}-4\right)\right\rceil \\
&=\left\lceil\frac{(9/4)^k-1}{(9/4)-1}\right\rceil\label{9/4}
\end{align}

Tokuda (1992) gap sequence begins with 
\begin{align}1,\, 4,\, 9,\, 20,\, 46,\, 103,\, 233,\, 525,\, 1182,\, 2660,\, 5985,\, 13467,\, 30301,\, 68178,\, 153401,\, 345152,\, 776591,\,...\end{align}

\clearpage By replacing the value of $\gamma_0=9/4=2.25$ by another $\gamma\in\mathbb{R}^+$, Tokuda gap sequence (1992) is generalised . Let $\gamma\in\mathbb{R}^+$. The $\gamma$-sequence is defined with the $k$-th increment $h_k$ given by, where $k\in\mathbb{N}_1$,
\begin{align}\label{Lee}
h_k=\left\lceil
\frac{\gamma^k-1}{\gamma-1}
\right\rceil
\end{align}

As inspired by the empirical approach of Ciura (2001) in studying gap sequences in Shellsort, the best value of $\gamma$, that minimises the number of comparisons on average, is approximated by a series of empirical experiments.

\section{Main Result}
A series of experiments in performing Shellsort for a fixed set of 10000 randomly generated permutations each consists of $N=10^{4+n}$ pairwise distinct elements, where $n=1,2,3,4$ represents the step of the experiment, is conducted. The value of $\gamma$ given by (\ref{result_gamma1}) empirically locally minimises the average number of comparisons in Shellsort, and having empirically less average number of comparisons than Tokuda (1992) gap sequence.\begin{align}\label{result_gamma1}
\gamma=2.243609061420001...
\end{align}

The first few increments of the $\gamma$-sequence given by (\ref{result_gamma1}) are
\begin{align}\label{result_sequence_1}
1,\, 4,\, 9,\, 20,\, 45,\, 102,\, 230,\, 516,\,1158,\,2599,\,5831,\,13082,\,29351,\,65853,\, 147748,\,331490,\,743735,\, ...\end{align}

\begin{figure}[h!]\begin{center}\begin{tikzpicture}
\begin{axis}[ table/col sep=comma, axis lines = left,
    width=\textwidth,
    height=0.40\textheight,
    axis x line=middle,
    xlabel = $N$,
    ylabel = Average Number of Comparisons / $\log_2{N!}$,
    xmode=log,ymode=normal,xmin=1,xmax=1e8,
    hide obscured x ticks=false,
    ytick = {1.0, 1.2, 1.4, 1.6, 1.8},
    yticklabel=\pgfkeys{/pgf/number format/.cd,fixed,precision=1,zerofill}\pgfmathprintnumber{\tick},
    ymin=1, ymax=1.8,
    legend pos=north west,
    legend cell align={left}] 
\addplot [color=black, dotted, mark=none] table [ x=nn, y=avga, ] {data3.csv};
\addplot [color=black, dashdotted,mark=none] table [ x=nn, y=avgc, ] {data3.csv};
\addplot [color=black, dashed, mark=none] table [ x=nn, y=avge, ] {data3.csv};
\addplot [color=black, mark=none] table [ x=nn, y=avgb, ] {data3.csv};
\legend{Tokuda (1992), Ciura (2001), $\gamma=2.376540775955158\dots$, $\gamma=2.243609061420001\dots$}
\end{axis}
\end{tikzpicture}\end{center}\caption{Average Numbers of Comparisons of the Gap Sequences for $N$ Pairwise Distinct Elements}\label{3seqcom}\end{figure}
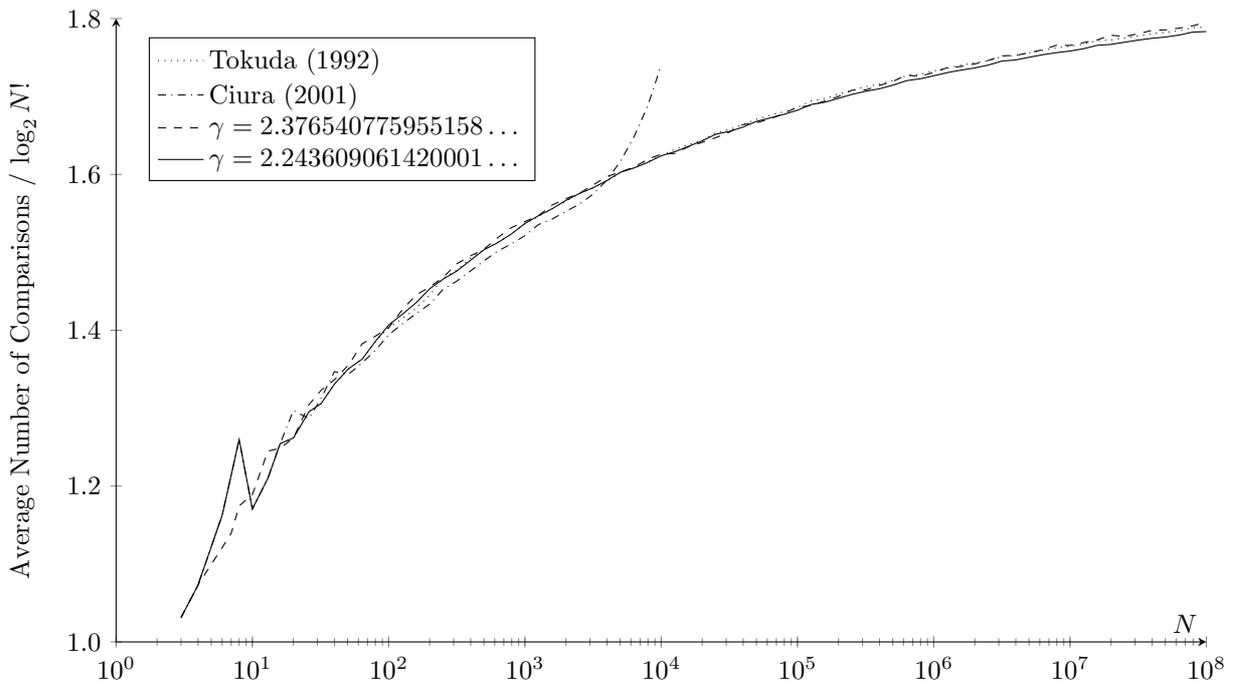

Figure \ref{3seqcom} summarise the result of the average numbers of comparisons for gap sequences of Tokuda (1992), Ciura (2001), the $\gamma$-sequences of (\ref{result_gamma1}) and (\ref{haha}), to slot a fixed set of 1000 randomly generated permutations each consists of $N=\lceil10^{k/10}\rceil$ pairwise distinct elements, where $k=1,...,80$. It is remarked that Ciura (2001) gap sequence is only intended to data of number of elements not more than 4000.

\vfill

\clearpage\section{Methodology}
Although as discovered, the optimal gap sequences in Shellsort, in the sense of minimised maximum number of comparisons, can consist of at least one increment that is larger than the half of the total number of the elements to sort (Lee, 2021), it seems to be a general practice to exclude all increments in the gap sequence that are larger than that half when the number of elements is large.

\begin{figure}[h]\begin{center}\begin{tikzpicture}
\begin{axis}[ table/col sep=comma, axis lines = left,
    width=\textwidth,
    height=0.34\textheight,
    axis x line=middle,
    xlabel = $N$,
    ylabel = Average Number of Comparisons / $\log_2{N!}$,
    hide obscured x ticks=false,
    xmode=log,ymode=normal,
    xmin=1e0,xmax=1e8,
    ymin=1, ymax=1.8,
    yticklabel style={/pgf/number format/.cd,fixed zerofill,precision=1},
    legend pos=north west,
    legend cell align={left},
]
\addplot [color=black, dashed, mark=none] table [ x=nn, y=avgd, ] {data3.csv};
\addplot [color=black, mark=none] table [ x=nn, y=avga, ] {data3.csv};
\legend{Including,Excluding}
\end{axis}
\end{tikzpicture}\end{center}\caption{Including or Excluding Increments of Magnitude Larger Than the Half of the Data Size}\label{N/2}\end{figure}
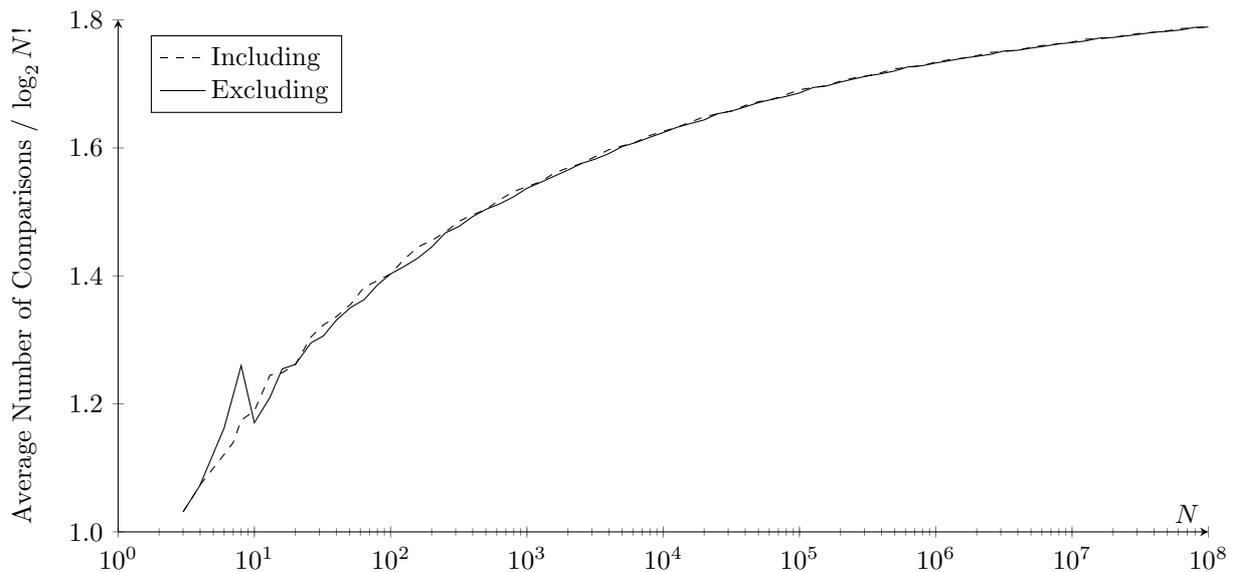

The convention is preserved in the study, with Figure \ref{N/2} illustrating the empirical benefit of the action, in applying the original Tokuda gap sequence in Shellsort to slot a fixed set of 1000 randomly generated permutations each consists of $N=\lceil10^{k/10}\rceil$ pairwise distinct elements, where $k=1,...,80$.

The range of the investigation is designated to be near the value of $\gamma_0=2.25$, and more precisely, \begin{align}\label{0C_step}
2.240000000000000 ... <\gamma\leq 2.260000000000000...
\end{align}

Notice that $(2.24,2.26]$ is uncountable and the set of all bounded gap sequences is countable, it suffices to consider the values of $\gamma$ in the boundary $\Gamma$ of a cylinder set,\begin{align}
\Gamma=\bigcup_{k\in\mathbb{N}_2}\left\{\gamma\in\mathbb{R}^+:\frac{\gamma^k-1}{\gamma-1}\in\mathbb{N}_1\right\}
\end{align}
The $\gamma$-sequence of $\gamma=2.24$ for $N=100000$, number of elements for $n=1$, is \begin{align}
1,\, 4,\, 9,\, 20,\, 45,\, 102,\, 228,\, 511,\, 1145,\, 2565,\, 5745,\, 12869,\, 28827, ...
\end{align}
The $\gamma$-sequence of $\gamma=2.26$ for $N=100000$ is \begin{align}
1,\, 4,\, 9,\, 20,\, 46,\, 105,\, 239,\, 540,\, 1220,\, 2758,\, 6235,\, 14090,\, 31845, ...
\end{align}

Notice that there are more than $31845-28827+1=3019$, the difference of the largest increments for $\gamma=2.24$ and $\gamma=2.26$, $\gamma$-sequences in step $n=1$, as it is possible that two $\gamma$-sequences having the same largest increment and differ in a smaller increment, such as $\gamma=2.240008844939135...$ and $\gamma=2.240011109091923...$.

\clearpage\section{Finding Suitable Range of $\gamma$ (Step $n=1$)}
For $n=1$, there are in total 5042 unique $\gamma$-sequences up to $N/2=50000$ satisfy (\ref{0C_step}).

\begin{table}[h!]\centering\begin{tabular}{>{\centering\arraybackslash}p{4.5cm}|>{\centering\arraybackslash}p{11.5cm}}
Boundary $\gamma\in\Gamma$  & $\gamma$-sequence up to $N/2=50000$ \\ \hline
 $\gamma=2.240004167478329...$
 & 1, 4, 9, 20, 45, 102, 228, 511, 1145, 2565, 5745, 12869, 28827 \\
 $\gamma=2.240008844939135...$ 
 & 1, 4, 9, 20, 45, 102, 228, 511, 1145, 2565, 5745, 12869, 28828 \\
 $\gamma=2.240011109091923...$ 
 & 1, 4, 9, 20, 45, 102, 228, 511, 1145, 2565, 5745, 12870, 28828 \\
 $\gamma=2.240018050483448...$
 & 1, 4, 9, 20, 45, 102, 228, 511, 1145, 2565, 5745, 12870, 28829 \\
 $\gamma=2.240024951543055...$ 
 & 1, 4, 9, 20, 45, 102, 228, 511, 1145, 2565, 5745, 12870, 28830 \\
 $\gamma=2.240024991652918...$ 
 & 1, 4, 9, 20, 45, 102, 228, 511, 1145, 2565, 5746, 12870, 28830 \\
  ... 
 & ... \\
 $\gamma=2.260002411823175...$ 
 & 1, 4, 9, 20, 46, 105, 239, 540, 1220, 2758, 6235, 14090, 31845
\end{tabular}
\caption{Partial List of the 5042 $\gamma$-Sequences for $N=10^5$}
\label{tab:List_step_1}
\end{table}

For reference, Tokuda (1992) gap sequence up to $N/2=50000$ is, one of the 5042 sequences and, \begin{align}\label{Tokuda,10^5}
1,\, 4,\, 9,\, 20,\, 46,\, 103,\, 233,\, 525,\, 1182,\, 2660,\, 5985,\, 13467,\, 30301\end{align}
 
Table \ref{1A_step} lists the top 8 $\gamma$-sequences that having the smallest empirical average numbers of comparisons.

\begin{table}[h!]\centering\begin{tabular}{>{\centering\arraybackslash}p{4.5cm}|>{\centering\arraybackslash}p{11.5cm}}
Boundary $\gamma\in\Gamma$  & $\gamma$-sequence up to $N/2=50000$ \\ \hline
 $\gamma= 2.243611492464884...$
 & 1, 4, 9, 20, 45, 102, 230, 516, 1158, 2599, 5831, 13082, 29351 \\
 $\gamma= 2.249132410615309...$ 
 & 1, 4, 9, 20, 46, 103, 233, 524, 1179, 2652, 5964, 13414, 30170 \\
 $\gamma= 2.243618431098797...$ 
 & 1, 4, 9, 20, 45, 102, 230, 516, 1158, 2599, 5831, 13082, 29353 \\
 $\gamma= 2.249198962980946...$
 & 1, 4, 9, 20, 46, 103, 233, 524, 1179, 2652, 5966, 13418, 30180 \\ 
 $\gamma= 2.243625146740100...$ 
 & 1, 4, 9, 20, 45, 102, 230, 516, 1158, 2599, 5831, 13083, 29353 \\
 $\gamma= 2.243693405249028...$ 
 & 1, 4, 9, 20, 45, 102, 230, 516, 1158, 2599, 5833, 13087, 29363 \\
 $\gamma= 2.243635250204080...$ 
 & 1, 4, 9, 20, 45, 102, 230, 516, 1158, 2599, 5831, 13083, 29355 \\
 $\gamma= 2.249138711873439...$ 
 & 1, 4, 9, 20, 46, 103, 233, 524, 1179, 2652, 5964, 13414, 30171 
\end{tabular}
\caption{Top 8 $\gamma$-Sequences for $N=10^5$}
\label{1A_step}
\end{table}

Most of the best $\gamma$-sequences have the first few increments being \begin{align}\label{1B_step}
1,\, 4,\, 9,\, 20,\, 45,\, 102,\, 230,\, 516,\,1158,\,2599,\,...
\end{align}

Hence, the best value of $\gamma$ can be restricted into the range that those $\gamma$-sequence satisfy (\ref{1B_step})
\begin{align}\label{1C_step}
2.243599966033135... < \gamma \leq 2.243705276469745... \end{align}

Since $\gamma_0=2.25$ does not satisfy (\ref{1C_step}), Tokuda (1992) gap sequence is not included into further searching steps. In other words, Tokuda (1992) gap sequence is not the empirically best gap sequence to sort $N=10^5$ pairwise distinct elements, and there are many $\gamma$-sequences perform a lot better than Tokuda (1992) gap sequence.

\clearpage\section{Reduced Range of $\gamma$ (Step $n=2$)}
For $n=2$, there are 375 unique $\gamma$-sequences up to $N/2=500000$ satisfy (\ref{1B_step}).

\begin{table}[h!]\centering\begin{tabular}{>{\centering\arraybackslash}p{4.5cm}|>{\centering\arraybackslash}p{11.5cm}}
Boundary $\gamma\in\Gamma$  & $\gamma$-sequence up to $N/2=500000$, continue from (\ref{1B_step}) \\ \hline
 $\gamma= 2.243600147031034...$
 & ..., 2599, 5830, 13081, 29350, 65850, 147741, 331471 \\
 $\gamma= 2.243600623830443...$ 
 & ..., 2599, 5830, 13081, 29350, 65850, 147741, 331472 \\
 $\gamma= 2.243601034054331...$ 
 & ..., 2599, 5830, 13081, 29350, 65850, 147741, 331473 \\
 $\gamma= 2.243601100628504...$
 & ..., 2599, 5830, 13081, 29350, 65850, 147742, 331473 \\
 $\gamma= 2.243601577425219...$ 
 & ..., 2599, 5830, 13081, 29350, 65850, 147742, 331474 \\
 $\gamma= 2.243601610816063...$ 
 & ..., 2599, 5830, 13081, 29350, 65850, 147742, 331475 \\
  ... 
 & ... \\
 $\gamma= 2.243705276469745...$ 
 & ..., 2599, 5833, 13088, 29365, 65887, 147832, 331692 \\
\end{tabular}
\caption{Partial List of the 375 $\gamma$-Sequences for $N=10^6$}
\label{tab:List_step_2}
\end{table}

For reference, Tokuda (1992) gap sequence up to $N/2=500000$ is, not one of the 375 sequences and, \begin{align}\label{Tokuda,10^6}
1,\, 4,\, 9,\, 20,\, 46,\, 103,\, 233,\, 525,\, 1182,\, 2660,\, 5985,\, 13467,\, 30301,\, 68178,\, 153401,\, 345152,\,...\end{align}

Table \ref{2A_step} lists the top 8 $\gamma$-sequences that having the smallest empirical average numbers of comparisons.

\begin{table}[h!]\centering\begin{tabular}{>{\centering\arraybackslash}p{4.5cm}|>{\centering\arraybackslash}p{11.5cm}}
Boundary $\gamma\in\Gamma$  & $\gamma$-sequence up to $N/2=500000$, continue from (\ref{1B_step}) \\ \hline
 $\gamma= 2.243609089542802...$
 & ..., 2599, 5831, 13082, 29351, 65853, 147748, 331490 \\
 $\gamma= 2.243608252437738...$ 
 & ..., 2599, 5831, 13082, 29351, 65853, 147748, 331488 \\
 $\gamma= 2.243607938780525...$
 & ..., 2599, 5831, 13082, 29351, 65852, 147747, 331488 \\
 $\gamma= 2.243607298880690...$ 
 & ..., 2599, 5831, 13082, 29351, 65852, 147747, 331486 \\
 $\gamma= 2.243606822100145...$ 
 & ..., 2599, 5831, 13082, 29351, 65852, 147747, 331485 \\
 $\gamma= 2.243698339269139...$ 
 & ..., 2599, 5831, 13082, 29351, 65853, 147748, 331489 \\
 $\gamma= 2.243608729214242...$ 
 & ..., 2599, 5833, 13087, 29364, 65884, 147823, 331671 \\
 $\gamma= 2.243697862745793...$ 
 & ..., 2599, 5833, 13087, 29364, 65885, 147826, 331676
\end{tabular}
\caption{Top 8 $\gamma$-Sequences for $N=10^6$}
\label{2A_step}
\end{table}

Most of the best $\gamma$-sequences have the first few increments being \begin{align}\label{2B_step}
1,\, 4,\, 9,\, 20,\, 45,\, 102,\, 230,\, 516,\,1158,\,2599,\,5831,\,13082,\,29351,\,...\end{align}

Hence, the best value of $\gamma$ can be restricted into the range that those $\gamma$-sequence satisfy (\ref{2B_step})
\begin{align} \label{2C_step}
2.243605292083476... < \gamma \leq 2.243611492464885... \end{align}

From Figure \ref{2.2436}, it is observed each $\gamma$-sequence that satisfies (\ref{2C_step}) performs empirically better than Tokuda (1992) gap sequence in (\ref{Tokuda,10^6}) to sort $N=10^6$ pairwise distinct elements. In other words, there are many gap sequences perform better than Tokuda (1992) gap sequence to sort larger number of elements.

\clearpage\vspace*{\fill}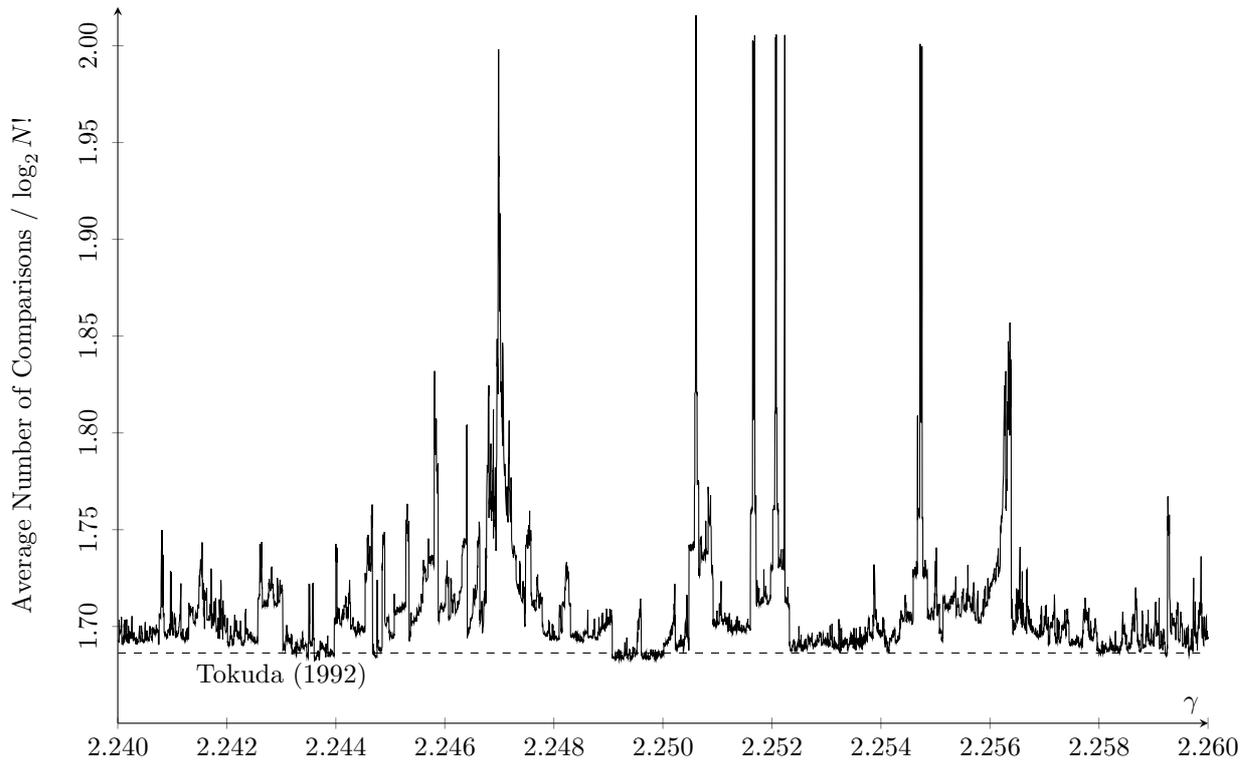
\begin{figure}[h]\begin{tikzpicture}
\centering
\begin{axis}[ axis lines = left,
    width=\textwidth,
    height=0.45\textheight,
    axis x line=middle,
    xlabel = \(\gamma\),
    ylabel = Average Number of Comparisons / $\log_2{N!}$,
    hide obscured x ticks=false,
    ytick = {1.70,1.75,1.80,1.85,1.90,1.95,2.00},
    yticklabel style={rotate=90,anchor=base,yshift=0.2cm},
    xmin=2.24, xmax=2.26,
    ymin=1.65, ymax=2.02,
    xticklabel style={/pgf/number format/.cd,fixed zerofill,precision=3},
    yticklabel style={/pgf/number format/.cd,fixed zerofill,precision=2}
]
\addplot[color=black, dashed, mark=none]
    coordinates {(2.24, 1.6862870322843)(2.26, 1.6862870322843)}
    node[pos=0.15,below] (end) {Tokuda (1992)};
\addplot+[mark=o,mark size=0pt,draw=black] table [x=gamma, y=mean, col sep=comma] {data_step1.csv} ;\end{axis}
\end{tikzpicture}\caption{Average Number of Comparisons of $\gamma$-sequence for $N=10^5$}\label{2.2500}\end{figure}
\vspace*{\fill}
\begin{figure}[h]\begin{tikzpicture}
\centering
\begin{axis}[ axis lines = left,
    width=\textwidth,
    height=0.45\textheight,
    axis x line=middle,
    xlabel = \(\gamma\),
    ylabel = Average Number of Comparisons / $\log_2{N!}$,
    xtick={2.24360,2.24362,2.24364,2.24366,2.24368,2.24370},
    ytick = {1.727,1.729,1.731,1.733,1.735,1.737,1.739},
    yticklabel style={rotate=90,anchor=base,yshift=0.2cm},
    xmin=2.243600147031034,
    xmax=2.243705276469745,
    ymin=1.726, ymax=1.74,
    xticklabel style={/pgf/number format/.cd,fixed zerofill,precision=5},
    yticklabel style={/pgf/number format/.cd,fixed zerofill,precision=3},
]
\addplot[color=black, dashed, mark=none]
    coordinates {(2.243599966033135, 1.7327300037751)(2.243705276469745, 1.7327300037751)}
    node[pos=0.15,below] (end) {Tokuda (1992)};
\addplot+[mark=o,mark size=0pt,draw=black] table [x=gamma, y=mean, col sep=comma] {data_step2.csv} ;\end{axis}
\end{tikzpicture}\caption{Average Number of Comparisons of $\gamma$-sequence for $N=10^6$}\label{2.2436}\end{figure}
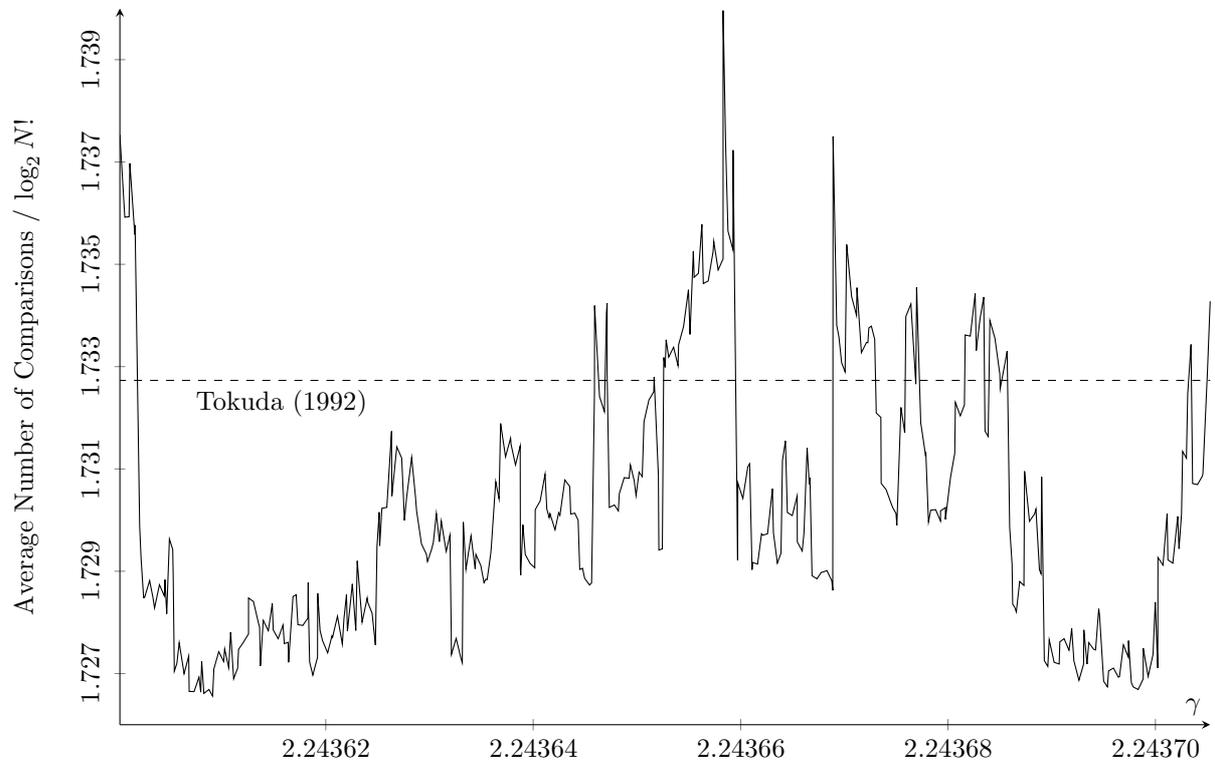\vspace*{\fill}

\clearpage\section{Approximation of $\gamma$ (Step $n=3$ and $n=4$)}
For $n=3$, there are 305 unique $\gamma$-sequences up to $N/2=5000000$ satisfy (\ref{2B_step}).

\begin{table}[h!]\centering\begin{tabular}{>{\centering\arraybackslash}p{4.5cm}|>{\centering\arraybackslash}p{11.5cm}}
Boundary $\gamma\in\Gamma$  & $\gamma$-sequence up to $N/2=5000000$, continue from (\ref{2B_step}) \\ \hline
 $\gamma= 2.243609062530634...$
 & ..., 29351, 65853, 147748, 331490, 743735, 1668650, 3743800 \\
 $\gamma= 2.243607900261593...$
 & ..., 29351, 65852, 147747, 331488, 743729, 1668636, 3743766 \\
 $\gamma= 2.243609053393101...$
 & ..., 29351, 65853, 147748, 331490, 743735, 1668650, 3743799 \\
 $\gamma= 2.243608216976799...$
 & ..., 29351, 65853, 147748, 331488, 743731, 1668640, 3743775 \\
 $\gamma= 2.243608232339401...$
 & ..., 29351, 65853, 147748, 331488, 743731, 1668640, 3743776 \\
 $\gamma= 2.243608729214242...$
 & ..., 29351, 65853, 147748, 331489, 743733, 1668646, 3743790 \\
 $\gamma= 2.243609089542802...$
 & ..., 29351, 65853, 147748, 331490, 743735, 1668651, 3743801 \\
 $\gamma= 2.243609018542523...$
 & ..., 29351, 65853, 147748, 331490, 743735, 1668650, 3743798
\end{tabular}
\caption{Top 8 $\gamma$-Sequences for $N=10^7$}
\label{3A_step}
\end{table}

Most of the best $\gamma$-sequences have the first few increments being \begin{align}\label{3B_step}
1,\, 4,\, 9,\, 20,\, 45,\, 102,\, 230,\, 516,\,1158,\,2599,\,5831,\,13082,\,29351,\,65853,\, 147748,\,331490,\,...\end{align}

Hence, the best value of $\gamma$ can be restricted into the range that those $\gamma$-sequence satisfy (\ref{3B_step})
\begin{align} \label{3C_step}
2.243608729214242... < \gamma \leq 2.243609089542802... \end{align}

For $n=4$, there are 238 unique $\gamma$-sequences up to $N/2=50000000$ satisfy (\ref{3B_step}).

\begin{table}[h!]\centering\begin{tabular}{>{\centering\arraybackslash}p{4.5cm}|>{\centering\arraybackslash}p{11.5cm}}
Boundary $\gamma\in\Gamma$  & $\gamma$-sequence up to $N/2=50000000$, continue from (\ref{3B_step}) \\ \hline
 $\gamma= 2.243609061420001 ...$
 & ..., 331490, 743735, 1668650, 3743800, 8399623, 18845471, 42281871 \\
 $\gamma= 2.243609060100173 ...$
 & ..., 331490, 743735, 1668650, 3743800, 8399623, 18845471, 42281870 \\
 $\gamma= 2.243609057472750 ...$
 & ..., 331490, 743735, 1668650, 3743800, 8399623, 18845471, 42281869 \\
 $\gamma= 2.243609049015995 ...$
 & ..., 331490, 743735, 1668650, 3743799, 8399623, 18845469, 42281866 \\
 $\gamma= 2.243609055217998 ...$
 & ..., 331490, 743735, 1668650, 3743800, 8399623, 18845470, 42281869 \\
 $\gamma= 2.243609062530634 ...$
 & ..., 331490, 743735, 1668650, 3743800, 8399623, 18845472, 42281871 \\
 $\gamma= 2.243609039080786 ...$
 & ..., 331490, 743735, 1668650, 3743799, 8399622, 18845468, 42281862 \\
 $\gamma= 2.243609042813992 ...$
 & ..., 331490, 743735, 1668650, 3743799, 8399622, 18845468, 42281864
\end{tabular}
\caption{Top 8 $\gamma$-Sequences for $N=10^8$}
\label{4A_step}
\end{table}

Most of the best $\gamma$-sequences have the first few increments being \begin{multline}\label{4B_step}
1,\, 4,\, 9,\, 20,\, 45,\, 102,\, 230,\, 516,\,1158,\,2599,\,5831,\,13082,\,29351,\,65853,\, 147748,\,\\
331490,\, 743735,\, 1668650,\, 3743800,\, 8399623,\, 18845471,\,... \end{multline}

Hence, the best value of $\gamma$ can be restricted into the range that those $\gamma$-sequence satisfy (\ref{4B_step})
\begin{align} \label{4C_step}
2.243609055217999... < \gamma \leq 2.243609061420001... \end{align}

\clearpage\vspace*{\fill}
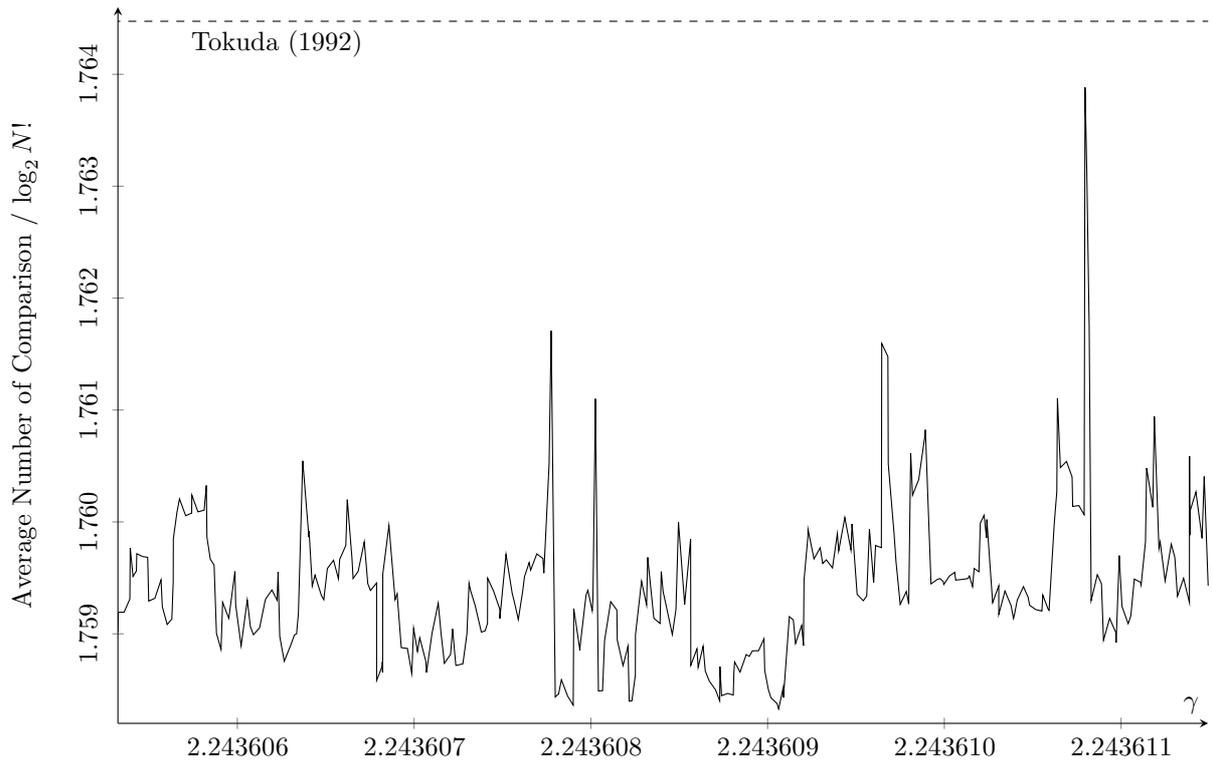
\begin{figure}[h]\begin{tikzpicture}
\centering
\begin{axis}[ axis lines = left,
    width=\textwidth,
    height=0.45\textheight,
    axis x line=middle,
    xlabel = \(\gamma\),
    ylabel = Average Number of Comparison / $\log_2{N!}$,
    xtick = {0.6, 0.7, 0.8, 0.9, 1.0, 1.1 },
    xticklabels = {2.243606, 2.243607, 2.243608, 2.243609, 2.243610, 2.243611},
    yticklabel style={rotate=90,anchor=base,yshift=0.2cm},
    ytick ={1.759, 1.760, 1.761, 1.762, 1.763, 1.764},
    xmin=0.532433128297427,
    xmax=1.149246488418854,
    ymin=1.7582, ymax=1.7646,
    xticklabel style={/pgf/number format/.cd,fixed zerofill,precision=5},
    yticklabel style={/pgf/number format/.cd,fixed zerofill,precision=3},
]
\addplot[color=black, dashed, mark=none]
    coordinates {(0.5292083476, 1.764473713543567)(1.149246488418854, 1.764473713543567)}
    node[pos=0.15,below] (end) {Tokuda (1992)};
\addplot+[mark=o,mark size=0pt,draw=black] table [x=gamma, y=mean, col sep=comma] {data_step3.csv} ;
\end{axis}
\end{tikzpicture}\caption{Average Number of Comparisons of $\gamma$-sequence for $N=10^7$}\label{2.2436}\end{figure}\vspace*{\fill}
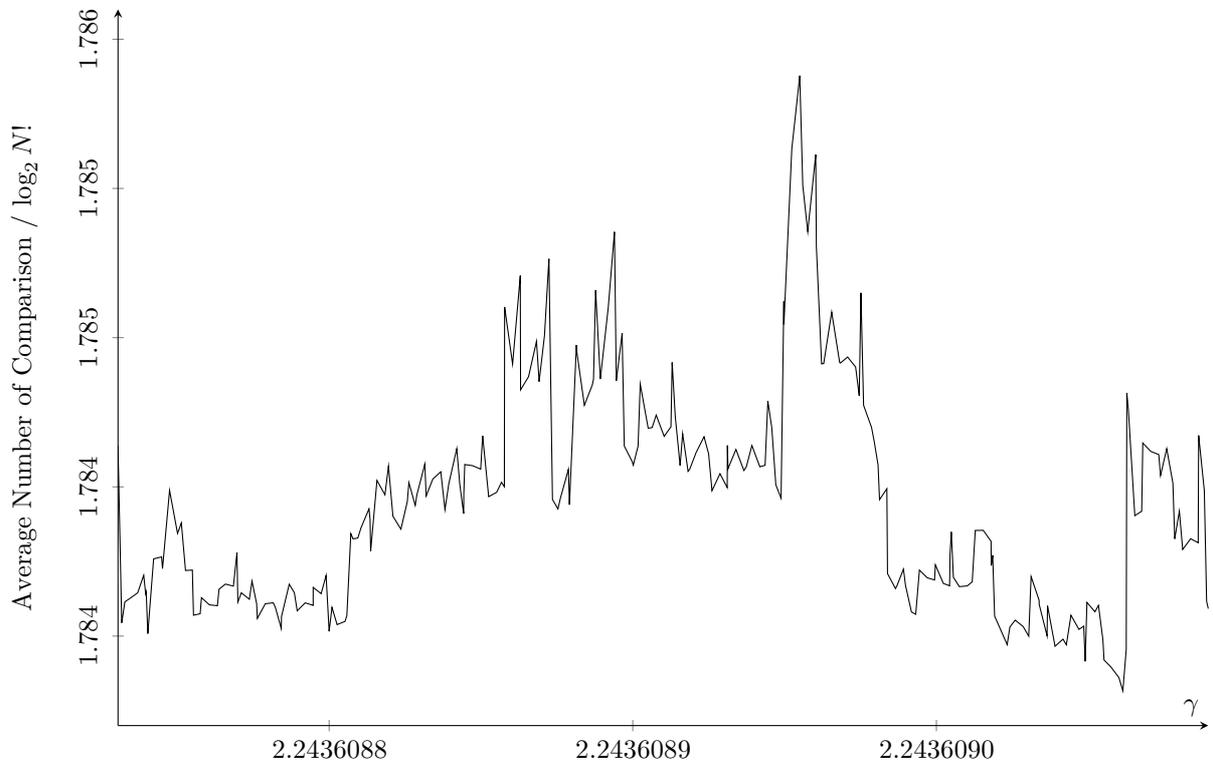
\begin{figure}[h]\begin{tikzpicture}
\centering
\begin{axis}[ axis lines = left,
    width=\textwidth,
    height=0.45\textheight,
    axis x line=middle,
    xlabel = \(\gamma\),
    ylabel = Average Number of Comparison / $\log_2{N!}$,
    xtick = {0.88, 0.89, 0.90, 0.91},
    xticklabels = {2.2436088, 2.2436089, 2.2436090, 2.2436091, 2.2436094},
    yticklabel style={rotate=90,anchor=base,yshift=0.2cm},
    ytick ={1.7835,1.7840,1.7845,1.7850,1.7855},
    xmin=0.8730454704,
    xmax=0.9089542802,
    ymin=1.7832, ymax=1.7856,
    xticklabel style={/pgf/number format/.cd,fixed zerofill,precision=5},
    yticklabel style={/pgf/number format/.cd,fixed zerofill,precision=3},
]
\addplot+[mark=o,mark size=0pt,draw=black] table [x=gamma, y=mean, col sep=comma] {data_step4.csv};
\end{axis}
\end{tikzpicture}\caption{Average Number of Comparisons of $\gamma$-sequence for $N=10^8$}\label{2.2436}\end{figure}\vspace*{\fill}

\clearpage\section{Parametrisation of Ciura (2001) Gap Sequence}
The value of $\gamma$ that makes Ciura (2001) gap sequence into a $\gamma$-sequence for each increment is gradually non-monotonically decreasing and has empty intersection. In particular, it is not a $\gamma$-sequence.

\begin{table}[h!]
\centering
\begin{tabular}{>{\centering\arraybackslash}p{2cm}|>{\centering\arraybackslash}p{9cm}}
Increment & Range of $\gamma$ \\ \hline
      1 & $\gamma\in\mathbb{R}^+$ \\
      4 & $2.000000000000000...\leq\gamma\leq 3.000000000000000...$ \\
     10 & $2.372281323269014...<\gamma\leq 2.541381265149110...$ \\
     23 & $2.356703176896190...<\gamma\leq 2.400693070226478...$ \\
     57 & $2.397935820913196...<\gamma\leq 2.410635279774417...$ \\
    132 & $2.380107890671399...<\gamma\leq2.384311537587710...$ \\
    301 & $2.361381573617817...<\gamma\leq2.362869501873037...$ \\
    701 & $2.356357448679420...<\gamma\leq2.356893893788949...$
\end{tabular}
\caption{The Values of $\gamma$ for Ciura (2001) Gap Sequence}
\label{tab:CiuraGamma}
\end{table}

Figure \ref{2.4500} summarise the results of the average numbers of comparisons for the $\gamma$-sequences, up to 1000 or 2000, where $\gamma\in(2.30,2.40)$ as near the range for the last increment in Ciura (2001) gap sequence, to sort 100000 fixed randomly generated permutations each consists of $N=4000$ pairwise distinct elements.

\vfill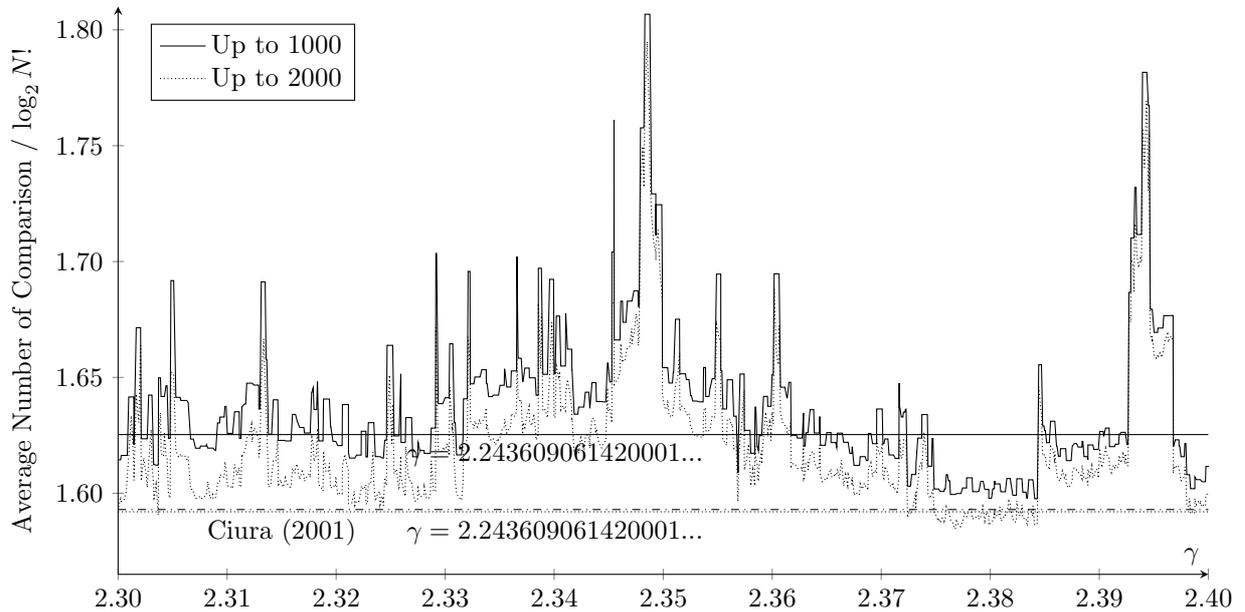
\begin{figure}[h!]\begin{tikzpicture}
\centering
\begin{axis}[ axis lines = left,
    width=\textwidth,
    height=0.37\textheight,
    axis x line=middle,
    xlabel = \(\gamma\),
    ylabel = Average Number of Comparison / $\log_2{N!}$,
    hide obscured x ticks=false,
    xmin=2.3, xmax=2.40,
    ymin=1.565, ymax=1.81,
    xtick = {2.30,2.31,2.32,2.33,2.34,2.35,2.36,2.37,2.38,2.39,2.40},
    xticklabel style={/pgf/number format/.cd,fixed zerofill,precision=2},
    yticklabel style={/pgf/number format/.cd,fixed zerofill,precision=2},
    legend pos=north west,
]
\addplot+[mark=o,mark size=0pt,draw=black] table [x=gamma, y=mean1, col sep=comma] {data9.csv};
\addplot+[mark=o,mark size=0pt,draw=black, densely dotted] table [x=gamma, y=mean2, col sep=comma] {data9.csv};
\addplot[color=black, mark=none]
    coordinates {(2.30, 1.625312734319742)(2.40, 1.625312734319742)}
    node[pos=0.4,below] (end) {$\gamma=2.243609061420001...$};
\addplot[color=black, densely dotted, mark=none]
    coordinates {(2.30, 1.591997793250734)(2.40, 1.591997793250734)}
    node[pos=0.4,below] (end) {$\gamma=2.243609061420001...$};
\addplot[color=black, dashdotted, mark=none]
    coordinates {(2.30, 1.592963670541602)(2.40, 1.592963670541602)}
    node[pos=0.15,below] (end) {Ciura (2001)};
\legend{Up to 1000, Up to 2000}
\end{axis}
\end{tikzpicture}\caption{Average Number of Comparisons of $\gamma$-sequence for $N=4000$}\label{2.4500}\end{figure}\vfill

It yields the best $\gamma$-sequence in this search up to 1000 (or 2000), \begin{align}\label{Ciura_sequence}
1,\, 4,\, 10,\, 23,\, 55,\, 131,\, 311,\, 739(,\, 1758)
\end{align}

It does not perform better than Ciura (2001) gap sequence, and the range of $\gamma$ that gives (\ref{Ciura_sequence}) is  \begin{align}\label{Ciura_gamma}
2.376288398428537...<\gamma\leq2.376800714745642...
\end{align}

From Figure \ref{2.4500}, it seems to suggest the existence of a better value of $\gamma\in(2.37,2.39)$ than (\ref{4C_step}) . Repeat the similar steps of the previous sections for $N=4\times 10^{3+n}$, the number of elements, where $n=1,2,3,4$ represents the step, the best $\gamma$-sequence has the first few increments being \begin{multline}\label{Con21}
1,\, 4,\, 10,\, 23,\, 55,\, 131,\, 311,\, 739,\, 1757,\,4175,\,9922,\,23580,\,56040,\,133182,\, 316512,\,\\
752203,\,1787642,\,4248404,\,... \end{multline}
The range of the values of $\gamma$ that the $\gamma$-sequence satisfies (\ref{Con21}) is
\begin{align} \label{4C_step}
2.376540775408273... < \gamma \leq 2.376540809782850... \end{align}
The minimum of average number of comparison achieves at \begin{align}\label{haha}
\gamma=2.376540775955158...\end{align}

However, from Figure \ref{3seqcom}, it returns out that $\gamma$-sequence given by (\ref{haha}) is not better than  given by (\ref{result_gamma1}) for large number of $N$. In other words, Ciura (2001) gap sequence cannot be improved by this method.

\vfill\section{Conclusion}

The empirically best value of $\gamma$ for large $N$, the number of elements to sort, approximated as
\begin{align}\label{con_gamma}
\gamma=2.243609061420001...\end{align}
, which is sufficient for most practical applications due to the limitation of double-precision arithmetic. 

On the other hand, many of the graphs for the average number of comparisons for $\gamma$-sequence against the value of $\gamma$ in a closed and bounded intervals look like fractals. Starting from the second graph, each is a locally zoom in graph of the previous one, and there is a certain kind of similarity in the sense that in each step of the search, as there are many local peaks and nadirs across the entire graphs.

For any $\gamma_1,\gamma_2\in\mathbb{R}^+$, if $1<\gamma_1<\gamma_2$ then there exists $K\in\mathbb{N}_1$ such that for any $k\in\mathbb{N}_1$, if $k>K$ then \begin{align}
\left\lceil\frac{\gamma_1^k-1}{\gamma_1-1}\right\rceil
<\left\lceil\frac{\gamma_2^k-1}{\gamma_2-1}\right\rceil
\end{align}
In particular, the function of assignment from $\gamma\in(1,+\infty)$ to the infinite $\gamma$-sequence is injective, and hence those graphs could be extended indefinitely for infinitely fine detail on arbitrarily small scales. To obtain precise images of those kind of fractals, the exact average number of comparisons for $\gamma$-sequence for any $\gamma$ in the desirable range is required. As the graph depends on $N$, a suitable scale for the vertical axis of the graph should be established such that the graph converges when $N$ tends to infinity.

\end{document}